\documentclass[10pt,reqno]{amsart}
\usepackage[applemac]{inputenc}
\usepackage[english]{babel}
\usepackage{graphicx}
\usepackage{hyperref}
\usepackage{amsmath,amsthm,amssymb,amscd}
\usepackage{cite}
\DeclareMathOperator{\Tr}{Tr}
\usepackage{color}

\newcommand{\rd}{{\mathrm d}}
\newcommand{\R}{\mathbb R}
\newcommand{\C}{\mathbb C}
\newcommand{\Z}{{\mathbb Z}}
\newcommand{\rdt}{\mathrm{d}^2}
\newcommand{\Os}{{S_{\mathrm{o}}}}
\newcommand{\e}{{\mathrm e}}
\newcommand{\Ccal}{{\mathcal C}}
\newcommand{\Lcal}{{\mathcal L}}
\newcommand{\Ncal}{{\mathcal N}}

\newcommand{\rhoout}{\rho_{\mathrm{out}}}
\newcommand{\rhoin}{\rho_{\mathrm{in}}}
\newcommand{\OS}{{S_{\mathrm{o}}}}
\newcommand{\MQFI}{{\mathcal M}_{\textrm{QFI}}}

\newcommand{\So}{OS\ }
\newcommand{\vertiii}[1]{{\left\vert\kern-0.25ex\left\vert\kern-0.25ex\left\vert #1 
    \right\vert\kern-0.25ex\right\vert\kern-0.25ex\right\vert}}

\begin{document}

\title[Measuring nonclassicality via operator ordering sensitivity]{Measuring nonclassicality of bosonic field quantum states via operator ordering sensitivity}

\author[S. De Bi\`evre, D. B. Horoshko, G. Patera, M.I. Kolobov]{Stephan De Bi\`evre$^{1}$, Dmitri B. Horoshko$^{2}$, Giuseppe Patera$^{2}$,\\ Mikhail I. Kolobov$^{2}$}

\address{
$^1$Univ. Lille, CNRS, UMR 8524 - Laboratoire Paul Painlev\'e; Equipe MEPHYSTO, 
INRIA, F-59000 Lille,
France\\
$^2$Univ. Lille, CNRS, UMR 8523 - PhLAM - Physique des Lasers Atomes et
Mol\'ecules, F-59000 Lille, France
}

\date{\today}

\begin{abstract}
We introduce a new distance-based measure for the nonclassicality of the states of a bosonic field, which outperforms the existing such measures in several ways. We define for that purpose the operator ordering sensitivity of the state which evaluates the sensitivity to operator ordering of the Renyi entropy of its quasi-probabilities and which measures the oscillations in its Wigner function. Through a sharp control on the operator ordering sensitivity of classical states we obtain a precise geometric image of their location in the density matrix space allowing us to introduce a distance-based measure of nonclassicality. We analyse the link between this nonclassicality measure and a recently introduced  quantum macroscopicity measure, showing how the two notions are distinct. 
\end{abstract}

\maketitle
\section{Introduction}
Questions arising in quantum information theory and in quantum chaos drive a continued interest in the exploration of the quantum-classical boundary.  
There is in this context a need for efficient criteria to determine the strength of the diverse nonclassical features of quantum states, such as: their Titulaer-Glauber nonclassicality~\cite{tigl65, hi85, balu86, hi87, hi89,  le91, agta92, le95, luba95, domamawu00, mamasc02, rivo02, kezy04, ascari05, se06, zapabe07, vosp14, ryspagal15, spvo15, kistpl16, al17, na17,  ryspvo17},  their degree of coherence and macroscopic nature~\cite{le02, leje11, go11, leje11b, frdu12, tope13, yu13, segisa14, frsagi15, ousefrgias15, yave16, dara17},   their degree of entanglement in multi-partite systems, their entanglement potential for mono-partite systems~\cite{ascari05, kistpl16}, their semi-classical breaking times in quantum chaos~\cite{gu90, gobr03}, and the links between these notions. In this paper we investigate the nonclassicality question for systems described with  bosonic variables. The well established definition of a Titulaer-Glauber classical state in this context is that it is a statistical mixture of coherent states, or equivalently, that its Glauber-Sudarshan $P$-function defines a probability on phase space~\cite{tigl65}.  
Otherwise, it is \emph{nonclassical}. In this paper, the term ``nonclassical'' will always be used in this precise sense. The two main issues in this respect are the identification of nonclassicality witnesses, or criteria, that allow to establish
if a given state is nonclassical and the definition of quantitative measures of nonclassicality, that allow to say how nonclassical a state is.

Indeed, a direct analysis of the $P$-function is rarely feasible because for many states the $P$-function is neither theoretically, nor experimentally readily accessible.
Consequently, to test for nonclassicality, various sufficient and more easily verified criteria have been designed. Some
generalize the well-known quantum optics criteria such as the negativity of the Mandel parameter, which detects sub-Poissonian photon statistics, and of the degree of squeezing~\cite{agta92, rivo02, ryspagal15}. Others involve the negativity of the Wigner function~\cite{le91, luba95, kezy04}, the entanglement potential of the state~\cite{ascari05},  or the minimal number of coherent states in terms of which one can write it as a superposition (See \cite{kistpl16, ryspvo17} and references therein).  While they capture various aspects of nonclassicality, they do not furnish a  nonclassicality measure. An alternative approach is to use a distance between a given state and the set $\Ccal$ of all classical states  as a nonclassicality measure. This idea was pursued using the trace norm~\cite{balu86, hi87, na17}, 
  the Hilbert-Schmidt norm~\cite{domamawu00}, 
  and the Bures distance~\cite{mamasc02}. It has been argued however that  the resulting nonclassicality measure depends on the arbitrariness in the choice of norm~\cite{spvo15}. Also, computing these distances has been possible only in very few cases, and even determining useful bounds on the nonclassicality has proven difficult~\cite{ascari05, ryspvo17, na17}. 

We propose a distance-based measure of nonclassicality avoiding those drawbacks. We first construct a specifically adapted Hilbert norm on the density operators, whose square we refer to as the ``operator ordering sensitivity'' (OS)  of the state $\rho$ (see~\eqref{eq:wigner0criterium} and~\eqref{eq:Osrewrite}). It measures the sensitivity of the Renyi entropy of its quasi-probability distributions to operator ordering, an eminently nonclassical notion. We show the OS provides  a simple and efficient sufficient condition for nonclassicality (see~\eqref{eq:wigner0criterium})  that is also necessary for pure states. Furthermore, as the square of a norm, the \So induces a distance $\Ncal(\rho)$ from $\rho$ to the set $\Ccal$ of all classical states that we propose as a new measure of nonclassicality (see~\eqref{eq:nonclassdef}).

We will establish that  the OS of $\rho$ yields a good approximation of the nonclassicality distance $\mathcal N(\rho)$, that it captures the intuitive physical ideas underlying nonclassicality well, and that it can often be more easily determined than existing criteria.

Another feature of the quantum-classical boundary is  ``quantum macroscopicity'' which, loosely speaking, evaluates the degree to which a quantum state is the superposition of macroscopically distinct states. In absence of a generally agreed upon definition, various measures of quantum macroscopicity have been proposed~\cite{le02, leje11, leje11b, go11, segisa14, ousefrgias15, frsagi15, yave16}. We will compare  a proposal based on the quantum Fisher information to the nonclassicality measure $\Ncal(\rho)$ and explain the relation between  nonclassicality and quantum macroscopicity.

 
\section{Ordering sensitivity: a nonclassicality witness}
For ease of notation, we shall concentrate on one-dimensional systems, characterized by an annihilation-creation operator pair $a, a^\dagger$. We introduce the $s$-ordered quasi-probabilities $W_s(\alpha)$ of a state with density matrix $\rho$ following~\cite{cagl69b}.  Let
\begin{equation*}
\chi_s(\xi)=\exp\left(s\frac{|\xi|^2}2\right)\chi_0(\xi),\quad \chi_0(\xi)=\mathrm{Tr}\rho D(\xi),
\end{equation*}
where $D(\xi)=\exp(\xi a^\dagger-\xi^*a)$.
Then
\begin{eqnarray}\label{eq:Ws}
W_s(\alpha)&=&\frac1{\pi^{2}}\int\chi_s(\xi)\exp(\xi^*\alpha-\xi\alpha^*)\rdt\xi.
\end{eqnarray}
Here $W_0(\alpha)$ is the Wigner function of $\rho$, $W_1(\alpha)$ its Glauber-Sudarshan $P$-function and $W_{-1}(\alpha)$ its Husimi function. We refer to $\chi_s(\xi)$ as the characteristic function of $W_s(\alpha)$. Then
\begin{equation}\label{eq:backdiffeq}
\partial_s W_s(\alpha)=-\frac18\Delta_\alpha W_s(\alpha).
\end{equation}
Here $\alpha=\alpha_1+i\alpha_2\in\C$ and $\Delta_\alpha=\partial_{\alpha_1}^2+\partial_{\alpha_2}^2$, so that $W_s(\alpha)$ is a solution of a backward diffusion equation, in which $s$ plays the role of the time, a crucial observation for what follows. Since $\Tr\rho=1$, one has in particular
$
\int W_s(\alpha)\rd^2 \alpha =1,
$
which, together with the fact that $W_s(\alpha)$ is real valued but not necessarily nonnegative, explains the terminology ``quasi-probability distribution.''   

The Wigner  function of $\rho$ is a continuous and square integrable function~\cite{fo89}, meaning that 
$
\|W_0\|_2^2:=\int|W_0|^2(\alpha)\rd^2 \alpha <+\infty.
$
For $s<0$, $W_s(\alpha)$ still has this property and actually becomes a smooth function, since it is the solution of a backward diffusion equation. For $s>0$, on the other hand, $W_s(\alpha)$, and in particular $P$, may develop strong singularities. 
Each single $W_s(\alpha)$ contains all information about $\rho$, which can in principle be reconstructed from it~\cite{cagl69b}.

Following~\cite{tigl65, hi85, mawo95}, we say a quantum state with density matrix $\rho$ is classical when the $P$-function of $\rho$  is a probability on the phase space $\C$. 
Below, we first establish a sufficient condition for a state $\rho$ to be  nonclassical based on its \emph{ordering sensitivity} $\OS(\rho)$, which measures the variation in the quasi-probability distributions $W_s(\alpha)$ of the $\rho$, for $s$ close to $0$: see~\eqref{eq:wigner0criterium}.

 For that purpose, we introduce the \emph{$s$-ordered entropy} of $\rho$:
\begin{equation}
H(s,\rho)=-\ln(\pi\|W_s\|_2^2).
\end{equation}
 This terminology is motivated by the observation that, when $W_s\geq0$, it defines a bona fide probability. $H(s,\rho)$ is then the (second order) Renyi entropy of that probability, one of many possible measures of its uncertainty or unpredictability. A  strongly localized or concentrated probability distribution corresponds to a low degree of uncertainty and a strongly negative Renyi entropy. Conversely, when a probability distribution is very much spread out,  its Renyi entropy is large and positive. 
For $s=0$,  $
H(0,\rho)=-\ln \mathrm{Tr}\rho^2 \geq 0
$
has a direct physical interpretation: it is the logarithm of the  purity $\mathcal P=\mathrm{Tr}\rho^2$ of $\rho$, and is in fact the Renyi entropy of its eigenvalues. It reaches its minimal value $0$ for pure states. 
Note that, from~\eqref{eq:backdiffeq} one finds
\begin{equation}\label{eq:Hprimeexpression}
H'(s,\rho)=\frac14\frac{\langle W_{s}, \Delta W_{s}\rangle}{\parallel W_{s}\parallel_2^2}=-\frac14\frac{\parallel \nabla W_{s}\parallel_2^2}{\parallel W_{s}\parallel_2^2}\leq 0.
\end{equation}
Hence $H$ is a decreasing function of $s$, reflecting the fact that $W_s(\alpha)$ solves a backward diffusion equation leading to an increase in entropy backward in the ``time'' $s$  and a decrease forward in time. Taking a further derivative one easily sees $H''(\rho, s)\leq 0$ so that $H$ is concave in $s$.

Our main tool for the characterization of  the nonclassicality of quantum states is the following bound on $H'$.\\
{\bf Theorem.}   If $\rho$ is a classical state, then 
\begin{equation}\label{eq:necessaryclassical}
0\leq -(1-s)H'(s,\rho)\leq 1, \quad -1\leq s< 1.
\end{equation}
The proof is given in Appendix~A and relies on~\eqref{eq:backdiffeq}.
The upper bound $1$ is sharp, since one easily checks that, for coherent states, $H'(s,|\alpha\rangle\langle\alpha|)=(s-1)^{-1}$.  The lower bound follows from~\eqref{eq:Hprimeexpression}. By evaluating~\eqref{eq:necessaryclassical} at $s=0$, we infer the following sufficient condition for nonclassicality of $\rho$:
\begin{equation}\label{eq:wigner0criterium}
\Os(\rho):=-H'(0,\rho)>1\Rightarrow \rho\ \text{is nonclassical}.
\end{equation}
We call~$\Os(\rho)$ the \emph{ordering sensitivity (OS)} of $\rho$. 
It measures the change in the $s$-ordered entropy of $\rho$, and hence the change in $W_s(\alpha)$, as $s$ varies close to $s=0$. This terminology is justified because different values of $s$ correspond to different operator orderings in the quantization procedure~\cite{cagl69a, cagl69b}.  The condition $\Os(\rho)>1$ can hence be paraphrased by saying that the state $\rho$ is strongly ordering sensitive and~\eqref{eq:wigner0criterium} says this implies the state is not classical, in agreement with ``operator ordering'' as a typical quantum feature. This provides a first argument in favour of  $\Os(\rho)$ as a nonclassicality probe.

A second argument comes from the observation that $\nabla W_0$ probes the oscillations and short range structures of $W_0$, associated in particular with interference fringes and with the negativity of $W_0$. Hence,~\eqref{eq:Hprimeexpression} implies $\Os(\rho)$ provides a measure of such features. The normalization by $\|W_0\|^2$ ensures that it is the frequency of the oscillations rather than their amplitude that is measured. Since interference fringes are a hallmark of quantum mechanics, physical intuition suggests large values of $\Os(\rho)$ are associated to a strongly nonclassical nature of the state.   A contrario, if $\rho$ is a classical state then, setting $s=0$ in~\eqref{eq:necessaryclassical} and using~\eqref{eq:Hprimeexpression}, one finds $\Os(\rho)\leq1$ or
\begin{equation}\label{eq:puritycontrol}
\|\nabla W_0\|_2^2\leq \frac{4}{\pi}\Tr\rho^2.
\end{equation}
Hence, for classical states, these oscillations in the Wigner distribution are \emph{purity limited}.
The less pure a classical state, the smaller they are. 

Equation~\eqref{eq:Hprimeexpression} therefore links two quantum phenomena: the sensitivity of $W_s(\alpha)$ to the operator ordering parameter $s$ and the oscillations of $W_s(\alpha)$ at fixed $s$. The quantity $\|\nabla W_0\|^2/\|W_0\|^2$ has been used previously in quantum chaos studies~\cite{gu90, gobr03} and both $\|\nabla W_0\|^2/\|W_0\|^2$ and $\|\nabla W_0\|^2$ have been proposed as measures of  ``quantum macroscopicity"~\cite{leje11, go11, leje11b}, but their relevance for that latter purpose has been contested~\cite{yave16} (see Section~\ref{s:QM} and Appendix~\ref{s:appQM} for details.) Our results above reinterpret  $\|\nabla W_0\|^2/\|W_0\|^2$ as the \So of the state and show it provides a nonclassicality witness. We will see it forms the basis for the construction of a nonclassicality measure. 

Thirdly, we consider the behaviour of the \So when the system interacts with a thermal bath with mean photon number $\langle n\rangle$. 
We use a simple input-output model~\cite{ascari05, se06} that can alternatively be interpreted as the action of a beam splitter~\cite{hara13}. The system, initially in the state $\rho_{\mathrm{in}}$, ends up in the state $\rhoout$ after interaction with the bath, characterized by an efficiency $0\leq \lambda\leq 1$, where
$$
\chi_{{\mathrm{out},1}}(\xi)=\chi_{\mathrm{in},1}(\sqrt\lambda\xi)\exp(-(1-\lambda)\langle n\rangle |\xi|^2).
$$
As a result, with $\overline s=1+\lambda^{-1}((s-1)-2(1-\lambda)\langle n\rangle)\leq 1$,
$$
W_{s, \mathrm{out}}(\alpha)=\lambda^{-1}W_{\overline s, \mathrm{in}}(\frac{\alpha}{\sqrt\lambda}).
$$
It follows that
$
\Os(\rhoout)=-\lambda^{-1}H'(\overline s, \rhoin)$, with $\overline s=1-\lambda^{-1}(1+2(1-\lambda)\langle n\rangle)<0.
$
Since $-H'$ is a non-decreasing function of $s$, this yields $\Os(\rhoout)\leq \lambda^{-1}\Os(\rhoin)$. For $\lambda$ close to $1$ and $\langle n\rangle$ large enough, this shows that $\Os(\rhoout)$ is lower than or equal to $\Os(\rhoin)$.  More precisely, in the weak coupling limit $\lambda\to1$ and $(1-\lambda)\langle n\rangle \to \overline e$,  this yields
$$
\lim_{\lambda\to1}\Os(\rhoout)=-H'(-2\overline e, \rhoin)\leq \Os(\rhoin).
$$
Since noisy environments destroy the quantal nature of states, this is again compatible with the interpretation of $\Os(\rho)$ as an indicator of the level of nonclassicality of $\rho$, a point further developed below, see~\eqref{eq:nonclassdef}-\eqref{eq:encadrement}. In fact, the above equation shows that, the noisier the environment (large $\overline e$), the more it decreases $\Os(\rho)$ and hence the nonclassicality of the initial state.

A final argument in favour of the pertinence of $\Os(\rho)$ as a nonclassicality probe comes from the analysis of $\Os(\rho)$ for pure states $\rho=|\psi\rangle\langle\psi|$. In that case,~\eqref{eq:Osrewrite} below implies
\begin{eqnarray}\label{eq:purestateHprime}
\Os (\rho)&=&\big\langle (Q-\langle Q\rangle)^2\big\rangle +\big\langle (P-\langle P\rangle)^2\big\rangle\nonumber\\
&=&2\langle(a^\dagger -\langle a^\dagger\rangle)(a-\langle a\rangle)\rangle+1.
\end{eqnarray}
Here $Q=\frac1{\sqrt2}(a^\dagger +a)$, $P=\frac{i}{\sqrt2}(a^\dagger-a)$.  This shows that for pure states the ordering sensitivity $\Os(\rho)$ captures the intuitive idea that they are strongly nonclassical when they have a large uncertainty.
Indeed, in classical mechanics pure states are identified with points in phase space and as such display no uncertainty, whereas in quantum mechanics, pure states must have uncertainty, of which $\Os(\rho)$ is a natural measure. 
The uncertainty principle implies $\Os(\rho)$  is larger than or equal to $1$ if $\rho$ is pure; it is equal to one only if $|\psi\rangle$ is a coherent state. Equations~\eqref{eq:necessaryclassical} and~\eqref{eq:purestateHprime}  therefore provide an alternative proof of the known fact that the only pure classical states are the coherent states~\cite{hi85}. It follows that the condition $\Os(\rho)>1$ is both necessary and sufficient for the nonclassicality of pure states. We will now show how to use $\OS(\rho)$ to construct a nonclassicality measure, thereby extending these ideas to mixed states. 

\section{A new nonclassicality measure}
We have seen the ordering sensitivity $\Os$ provides a sufficient condition for nonclassicality which is also necessary for pure states. We now construct, using $\Os$, a nonclassicality measure for all states. For that purpose, we first interpret $\Os$ geometrically. 
We define, for two operators $A,B$ with finite trace,
 \begin{equation}\label{eq:innerproduct}
 \langle A, B\rangle =\frac12 \Tr\left([A^\dagger, Q][Q, B] +[A^\dagger, P][P, B]\right).
 \end{equation}
This expression is linear in $B$, anti-linear in $A$, and positive when $B=A$. We will set
$
\vertiii{A}=\langle A, A\rangle^{1/2}.
$
If $\vertiii{A}=0$, $A$ vanishes (see Appendix~\ref{s:appHS}). Hence the above expression defines an inner product. We write $\Lcal_{\textrm{HS}}^{(1)}$ for the corresponding Hilbert space of operators. One has (See~\cite{gu90} and Appendix~\ref{s:appHS})
\begin{equation}\label{eq:Osrewrite}
\Os(\rho)=-\frac12\frac{\Tr \left([Q,\rho]^2+[P,\rho]^2\right)}{\Tr \rho^2}=\vertiii{\tilde\rho}^2,
\end{equation}
where $\tilde\rho ={\rho}/{\sqrt{\Tr(\rho^2)}}$. This shows that the \So of $\rho$ is a norm. The map $\rho\to\tilde\rho$ is the normalization of $\rho$ for the Hilbert-Schmidt norm.  

We now reformulate~\eqref{eq:wigner0criterium}: $\rho\in\Ccal \Rightarrow \vertiii{\tilde\rho}\leq 1.$
In other words, $\tilde\Ccal$, which is the image of $\Ccal$ under the map $\rho\to\tilde\rho$, is contained inside the unit ball of $\Lcal_{\textrm{HS}}^{(1)}$. 
Conversely, when $\tilde\rho$ is outside this unit ball, $\rho$ is nonclassical. We define the distance from $\rho$ to $\Ccal$ by 
$
d(\rho, \Ccal)=\inf_{\sigma\in \Ccal}\vertiii{\tilde\rho-\tilde\sigma}
$
and propose it as a quantitative nonclassicality measure for $\rho$ by defining the \emph{nonclassicality} $\Ncal(\rho)$  of $\rho$ via
\begin{equation}\label{eq:nonclassdef}
\Ncal(\rho)=d(\rho, \Ccal).
\end{equation}
Note that it is a continuous function of $\rho$. Clearly, $\Ncal(\rho)>0$ implies $\rho$ nonclassical and $\rho$ classical implies $\Ncal(\rho)=0$ (see Appendix~\ref{s:appHS} for details). 

One could object that, since we have no good understanding of the precise shape of $\Ccal$, this distance cannot be readily computed, as for the distances previously introduced in the literature. However, since $\tilde \Ccal$ lies inside the unit ball, and since the \So of classical states can be arbitrarily small, the triangle inequality for norms implies (Appendix~\ref{s:appHS})
\begin{equation}\label{eq:encadrement}
\vertiii{\tilde \rho}-1\leq \Ncal(\rho) \leq \vertiii{\tilde \rho}.
\end{equation}
Hence, if $\vertiii{\tilde\rho}\gg1$, then $\vertiii{\tilde\rho}$ provides a very good estimate of $\Ncal(\rho)$. In addition,~\eqref{eq:Osrewrite} expresses $\Os(\rho)$ directly in terms of the density matrix $\rho$ itself, without referring to its quasi-probabilities $W_s(\alpha)$. This, as we will see, is a distinct advantage in its computation. Of course, when, for example through quantum tomography, the Wigner function of the state is experimentally accessible, then the \So can be determined directly from~\eqref{eq:Hprimeexpression} and hence the nonclassicality of the state assessed quantitatively.

\section{Computing the ordering sensitivity of pure and mixed states: examples}\label{s:examples}
\noindent\emph{Computing the ordering sensitivity: pure states.}
One finds, as anticipated above and using~\eqref{eq:purestateHprime}, that $\Os(|\alpha\rangle\langle\alpha|)=1$. For squeezed states $|\alpha, z\rangle, z=r\e^{i\varphi}$, one finds $\OS(|\alpha, z\rangle\langle\alpha, z|)=\cosh(2r)$: increased squeezing leads to increased nonclassicality. Also, $\Os(|n\rangle\langle n|)=2n+1$: the number states are increasingly far from the set $\Ccal$ of classical states as $n$ grows, corroborating their increasing nonclassicality. For the even/odd coherent states $|\psi_\pm\rangle \propto \left(|\alpha\rangle\pm |-\alpha\rangle\right)$ one finds 
$$
\Os\left(|\psi_\pm\rangle\langle \psi_\pm|\right)=2\langle a^\dagger a\rangle +1=2|\alpha|^2\frac{1\mp\langle\alpha|-\alpha\rangle}{1\pm\langle\alpha|-\alpha\rangle}+1.
$$
Hence their nonclassicality grows as $|\alpha|^2$; the same is true for $N$-component cat states introduced in~\cite{hara13, hodekopa16} (see Appendix~\ref{s:appCS}). 
In contrast, for such states, the Mandel parameter and the degree of squeezing, as well as the method of moments, provide inefficient nonclassicality witnesses when $\alpha$ is large (see Appendix~\ref{s:appCS}). 
The entanglement potential of the N-component cat states saturates at $\ln N$~\cite{ascari05, hodekopa16} for large $|\alpha|$ so that it does not capture the nonclassicality growth with growing $|\alpha|$.
Similarly, the degree of nonclassicality introduced in~\cite{vosp14, ryspvo17} equals $N$, independently of $\alpha$.
Finally, our approach here has an essential advantage over the one using the trace distance $\delta_\Ccal(\rho)=\frac12\inf_{\sigma\in\Ccal}\Tr\left(|\rho-\sigma|\right)$ as a measure of nonclassicality. Indeed, $\delta_\Ccal(|n\rangle\langle n|)$ tends to its maximal possible value $1$ as $n$ grows, whereas $\delta_\Ccal(|\psi_{\pm}\rangle\langle \psi_\pm|)$ saturates at $1/2$ for large $|\alpha|$~\cite{na17}. It is therefore insensitive to the increased phase space spread of those odd/even coherent states. In addition, to the best of our knowledge, $\delta_\Ccal(|\psi\rangle\langle\psi|)$  has not been computed for more complex pure states. In contrast, \eqref{eq:purestateHprime} shows $\Os(|\psi\rangle\langle\psi|)$ is readily determined from $\langle a\rangle$ and $\langle a^\dagger a\rangle$. 

\begin{figure}
\includegraphics[width=0.6\columnwidth]{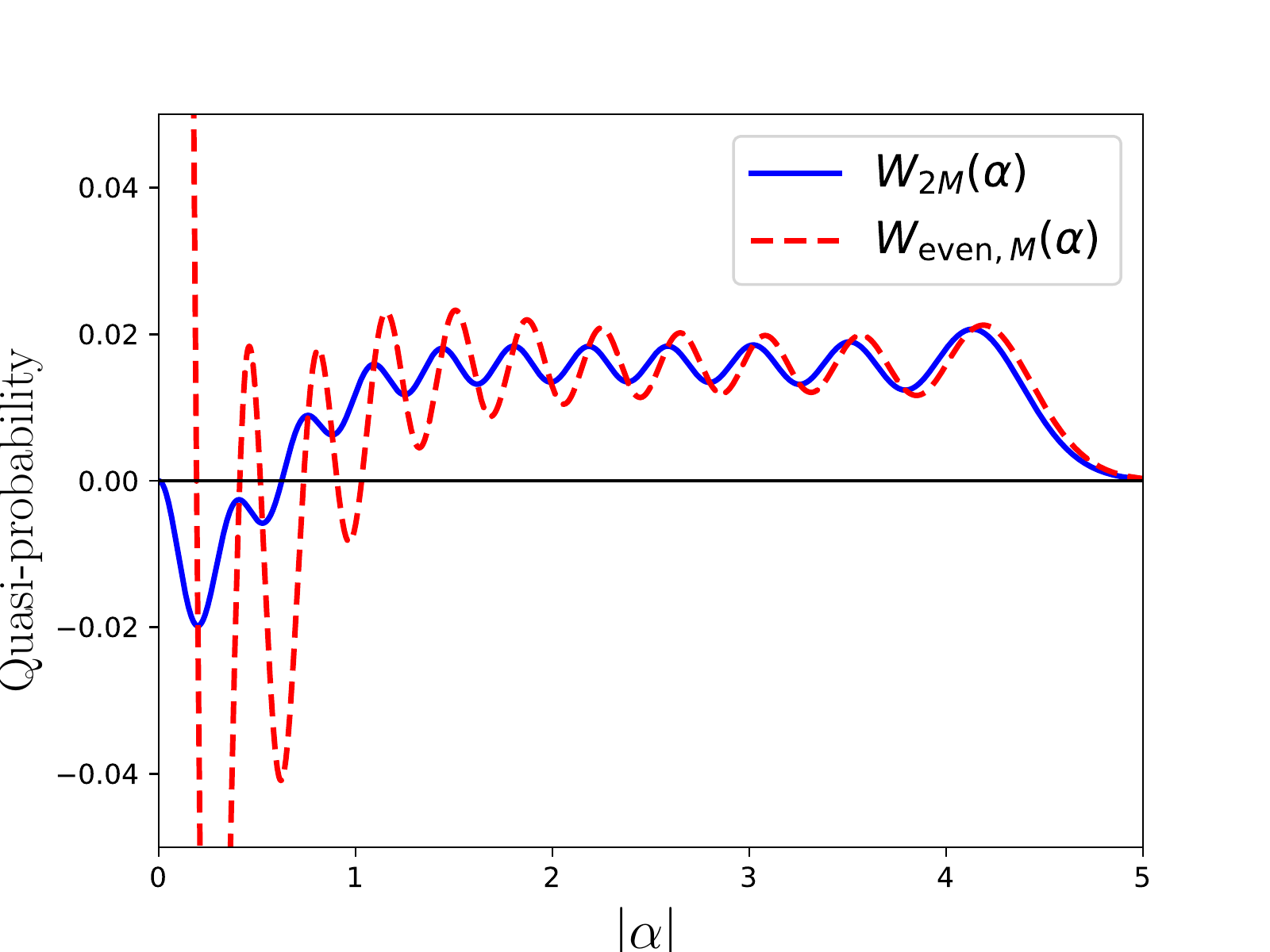}
\caption{Colour online. Plots of the Wigner functions of $\rho_{2M}$ (solid line) and of $\rho_{\mathrm{even}, M}$ (dashed line) as a function of $|\alpha|$ for $M=10$. Both states have comparable photon number: $\Tr (\rho_{2M}a^\dagger a)=M+\frac12$, $\Tr(\rho_{\mathrm{even}, M}a^\dagger a)=M+1$. The oscillations in the Wigner function of one are visibly much more pronounced than in the other, as detailed in the text.  \label{fig:wigner}}
\end{figure}

\noindent\emph{Computing the ordering sensitivity: mixed states.}  Let $\rho$
$
\rho=\sum_i p_i|i\rangle\langle i|
$, where  $\langle i|j\rangle=\delta_{ij}$. Then
$\langle\tilde\rho, \tilde\rho\rangle=\tilde p^T K\tilde p$, with, for $i\not=j$ (see Appendix~\ref{s:appHS})
\begin{equation}\label{eq:Kmatrix}
K_{ii}=\Delta Q_i^2+\Delta P_i^2, K_{ij}=-\left(|\langle i|Q|j\rangle|^2 +|\langle i|P|j\rangle|^2\right).
\end{equation}
The computation of $\Os(\rho)$ is therefore reduced to the computation of field quadratures in the eigenstates $|i\rangle$ of $\rho$ followed by the analytical or numerical computation of a matrix element of $K$.  In comparison, the computation of the trace, Hilbert-Schmidt or Bures distances has not been achieved for mixed states. Also,  the determination of the Mandel parameter and a fortiori the use of the moment method, require the computation of higher moments in $a, a^\dagger$ (see Appendix~\ref{s:appCS} for details). From~\eqref{eq:Kmatrix} it follows $\Os(\rho)$ is less than the weighted average $\sum_i \tilde p_i^2 (\Delta Q_i^2+\Delta P_i^2)$ of the ordering sensitivities of the basis states $|i\rangle$.  When the off-diagonal terms are small, this bound can be reached. For example, when the $|i\rangle$ are given by the number states $|n\rangle$, one has $K_{nn}=2n+1$ and $K_{n n+1}=- (n+1)$.   Considering $\rho_{\mathrm{even}}=\sum_n p_{2n}|2n\rangle\langle 2n|$ one then finds
$$
\Os(\rho_{\mathrm{even}})=\sum_n \tilde p_{2n}^2 (\Delta Q_{2n}^2+\Delta P_{2n}^2)=1+4\sum_n \tilde p_{2n}^2 n.
$$
When $p_0=0, p_2=\dots=p_{2M}=1/M$, this yields
$\Os(\rho_{\mathrm{even}, M})=1+2(M+1)$. These states are therefore increasingly nonclassical as $M$ grows and show strong oscillations in their Wigner function.  For $\rho_M=M^{-1}\sum_{n=1}^{M}|n\rangle\langle n|$~\cite{leje11, ryspvo17}, on the contrary, the off-diagonal elements reduce the \So substantially and  $\Os({\rho_M})=1+2M^{-1}$. The $\rho_M$ are only weakly nonclassical: they remain at a distance at most $1+M^{-1}$ from $\Ccal$ and their Wigner function shows only small fluctuations   (Fig.~\ref{fig:wigner}). On the other hand, the Mandel parameter grows for both $\rho_{\mathrm{even}, M}$ and $\rho_M$ as $M$  thereby failing to detect their nonclassicality for large $M$. Also, despite the fact that these two types of states are very different, the degree of nonclassicality introduced in~\cite{vosp14, ryspvo17} is $2M$ for both and does not distinguish them. 
Similar computations allow to determine the ordering sensitivity of a mixture of a thermal and a Fock state $|m\rangle$~\cite{leje11}, which is strongly nonclassical for large $m$, as well as for truncated thermal states~\cite{le91} and for (single) photon added thermal~\cite{agta92, zapabe07} states, which are found to be weakly nonclassical with ordering sensitivities between $0.5$ and $2$, depending on the temperature~\cite{future}. Note that the nonclassicality of the above states cannot be revealed through squeezing since they are phase-insensitive.

\section{Nonclassicality versus quantum macroscopicity}\label{s:QM}
In contrast to the ``nonclassicality'' of a state,  there is no generally agreed upon definition of its ``quantum macroscopicity'',  a property for which a variety of measures have been proposed recently~\cite{le02, leje11, go11, frdu12, segisa14, ousefrgias15, frsagi15, yave16}.  One such measure~\cite{ousefrgias15, yave16} uses the quantum Fisher information $\mathcal F(\rho, Q_\theta)$ of the quadratures $Q_\theta=Q\cos\theta + P\sin\theta$: 
$$
\MQFI(\rho)=\frac14\max_\theta \mathcal F(\rho,Q_\theta).
$$
We show in Appendix~\ref{s:appQM} that 
\begin{equation}\label{eq:mqfiwitness}
\MQFI(\rho)>\frac12\Rightarrow \rho \ \textrm{nonclassical},
\end{equation}
proving that $\MQFI$ is a nonclassicality witness, as $\OS$: a large value of $\MQFI$ guarantees the state is Glauber-Titulaer nonclassical.  It is natural to ask how $\MQFI$ and $\OS$ are related. One may notice that on many states they behave similarly. Indeed, on thermal states and on $\rho_M$ (see the examples above), $\OS=2\MQFI$, so that they coincide except for a normalization (see Appendix~\ref{s:appQM}). The two can also be very different. For example, $\MQFI$ is a  less efficient nonclassicality witness than $\OS$ for truncated vacuum states, while it is more efficient than $\OS$ for squeezed thermal states. However, and more importantly, a large $\MQFI$ does not imply a large $\Ncal$, as the simple example $\rho_k=(1-\frac{M_*}k)|0\rangle\langle 0|+\frac{M_*}k|k\rangle\langle k|$ shows: $\MQFI(\rho_k)=\frac12+M_*$, $\Ncal(\rho_k)\to0$ when $k\gg1$. So, when $M_*$ is large, $\MQFI$ is large, while the nonclassicality $\Ncal$ remains small (see Appendix~\ref{s:appQM} for details). Consequently, if $\MQFI$ does indeed correctly capture the idea of ``quantum macroscopicity'', as proposed in~\cite{ousefrgias15, yave16}, then large quantum macroscopicity does not imply large nonclassicality in the sense of Glauber-Titulaer. This would seem to indicate that, even for a single mode, $\MQFI$ captures ``macroscopic'' and/or ``quantum'' features of such states that are different from the nonclassicality associated with a nonpositive Sudarshan-Glauber $P$ function and revealed by their OS. What these features are, remains unclear. In all examples we treated, a large nonclassicality $\Ncal$ implies a large $\MQFI$, but whether this is generally true is not clear.

 
\section{Conclusions}
We have constructed a new measure $\Ncal(\rho)$ for the nonclassicality of the states of a single-component boson field. $\Ncal(\rho)$ is a distance to the set $\Ccal$ of all classical states, defined in terms of the ordering sensitivity (OS) of the state, a new entropic notion that we introduced and that evaluates the sensitivity of the state to operator ordering. We have proven the crucial properties that all classical states have an \So less than one 
and that, 
when the \So of a density matrix is large, it provides a good approximation of $\Ncal(\rho)$. The 
\So is easily computable in terms of field quadratures, captures several intuitive features of nonclassicality 
naturally, and detects in many cases nonclassicality more efficiently than previously known indicators. We have finally compared the nonclassicality $\Ncal(\rho)$ to a recent proposal for the measure of ``quantum macroscopicity'' based on the Quantum Fisher Information.

We expect that the extension of the ideas developed here to multi-mode fields
will help to clarify the relations between the Titulaer-Glauber nonclassicality and other manifestations of nonclassical behaviour such as entanglement. 
Finally, since the ordering sensitivity $\OS(\rho)$  is defined in terms of the $s$-ordered entropy $H(s,\rho)$, it can be used to establish fundamental relations between quantum theory and thermodynamics~\cite{future}.

\vskip 0.5cm
\noindent\textbf{Acknowledgments.}
This work was supported in part by the Labex CEMPI (ANR-11-LABX-0007-01) and by the Nord-Pas de Calais Regional Council and FEDER through the Contrat de Projets \'Etat-R\'egion (CPER), and in part by the European Union'Äôs Horizon 2020 research and innovation programme under grant agreement No 665148 (QCUMbER).

\appendix
\section{Proof of the bound~(5) on the ordering sensitivity of classical states}
We need to show that, if $\rho$ is a classical state, then
$0\leq -(1-s)H'(s,\rho)\leq 1, s< 1$.
We will use equation~\eqref{eq:backdiffeq}. If $\rho$ is classical, it is of the form
$
\rho=\int_{\C} |\alpha\rangle\langle\alpha| P(\alpha)\rd \alpha,
$
where $P(\alpha)\rd\alpha$ is a probability measure. Hence, $W_s$ solves for $s<1$ a backward diffusion equation with initial condition $W_1\rd\alpha=P\rd\alpha$, so that, for $0\leq \tau=1-s$, $D=\frac14$,
\begin{equation}
W_{1-\tau}(\alpha)=\int G_\tau(\alpha-\alpha')P(\alpha')\rd\alpha',\quad \text{with}\quad G_\tau(\alpha)=(2\pi D\tau)^{-1}\exp\left(-\frac{|\alpha|^2}{2D\tau}\right).\nonumber
\end{equation}
Next, recalling that $H(s)=-\ln (\pi \parallel W_s\parallel_2^2)$, one sees from equation~\eqref{eq:backdiffeq} that for $\tau>0$
$$
-H'(1-\tau)=\frac{-\frac{\rd}{\rd \tau}\parallel W_{1-\tau}\parallel_2^2}{\parallel W_{1-\tau}\parallel_2^2}=\frac14\frac{\parallel \nabla W_{1-\tau}\parallel_2^2}{\parallel W_{1-\tau}\parallel_2^2}\geq 0.
$$
Hence, proving equation~\eqref{eq:necessaryclassical} is equivalent to proving,
for all $0< \tau$,
$$
\Lambda(\tau):=-\partial_\tau\parallel W_{1-\tau}\parallel_2^2- \frac1\tau\parallel W_{1-\tau}\parallel_2^2\leq 0.
$$
Then
\begin{eqnarray*}
0\leq-\partial_\tau \|W_{1-\tau}\|^2&=&-2\int W_{1-\tau}(\alpha)\partial_\tau W_{1-\tau}(\alpha)\rd\alpha,\\
&=&-2\int\int W_{1-\tau}(\alpha)\left(-\frac1\tau +\frac{|\alpha-\alpha_2|^2}{2D\tau^2}\right)G_\tau(\alpha-\alpha_2)W_1(\alpha_2)\rd\alpha\rd\alpha_2\\
&=&2\int\int W_{1}(\alpha_1)G_\tau(\alpha-\alpha_1)\tilde G_\tau(\alpha-\alpha_2)W_{1}(\alpha_2)\rd \alpha\rd\alpha_1\rd\alpha_2,
\end{eqnarray*}
where
$
\tilde G_\tau(\alpha)=-\left(-\frac1\tau +\frac{|\alpha|^2}{2D\tau^2}\right)G_\tau(\alpha).
$
So
$$
\Lambda(\tau)=\int W_1(\alpha_1)S_\tau(\alpha_1, \alpha_2)W_1(\alpha_2)\rd \alpha_1\rd\alpha_2
$$
with
$$
S_\tau(\alpha_1, \alpha_2)=\int G_\tau(\alpha-\alpha_1)\left[\frac2{\tau}-\frac1\tau-\frac{2|\alpha-\alpha_1|^2}{2D\tau^2}\right]G_\tau(\alpha-\alpha_2)\rd\alpha.
$$
Integrating over $\alpha$ in this expression, one finds
$$
S_\tau(\alpha_1, \alpha_2)=\frac1{4\pi D \tau} \exp(-\frac{|\alpha_1-\alpha_2|^2}{4D\tau})\left[-\frac{|\alpha_1-\alpha_2|^2}{4D\tau^2}\right].
$$
This expression is negative, so it follows from the positivity of $W_1(\alpha)\rd \alpha$ for all $\alpha$ that $\Lambda(\tau)\leq 0$, which is the desired result. Note that $\Lambda(\tau)=0$ if and only if $W_1(\alpha)\rd\alpha$ is a Dirac delta measure, which corresponds to a coherent state. 

\noindent{\bf Remarks.} (a) In the presence of $d$ modes, the same computation as above yields
$$
 0\leq -(1-s)H'(s,\rho)\leq \frac{d}{2}, \quad s< 1.
 $$
The theory can therefore be adapted to multi-component bose fields.
(b) We point out the following subtlety. That $P(\alpha)\rd \alpha$ is a probability measure does not imply $P$ is square integrable: $\int P^2(\alpha)\rd\alpha$ may be infinity, such as when $P(\alpha)=\delta(\alpha-\alpha_0)$. However, the Young inequality for convolutions~\cite{ma95} guarantees that $W_s$ for $s<1$ is square integrable. This is not true for all states, but it is for classical states and legitimizes the computations of the $L^2$-norms above.

\section{Ordering sensitivity as a Hilbert space norm}\label{s:appHS}
We first show~\eqref{eq:Osrewrite}. An easy computation shows that, if $W_A$ is the Wigner function of an operator $A$, then $\partial_{\alpha_1}W_A$ is the Wigner function of $\sqrt2 i[A,P]$ and $\partial_{\alpha_2}W_A$  of $\sqrt2 i[A,Q]$.  It follows then from~\eqref{eq:innerproduct} that
\begin{equation} \label{eq:wigner_sobolev}
 \langle A, B\rangle  = \frac{\pi}{4}\int \overline{\nabla W_A(\alpha)}\nabla W_B(\alpha)\rd \alpha.
 \end{equation}
Taking $A=B=\rho$, one finds~\eqref{eq:Osrewrite} . Note that what precedes is formal, since, as a result of the fact that $Q$ and $P$ are unbounded operators, $\vertiii{A}$ can be equal to $+\infty$, and the inner product $\langle A, B\rangle$ may be ill defined. To avoid this problem, we proceed as follows. We write $\Lcal_{\textrm{HS}}$ for the set of Hilbert-Schmidt operators, \emph{i.e.} those operators $A$ for which $\Tr(A^\dagger A)<+\infty$. We then define the following vector space of operators:
$$
 \Lcal_{\textrm{HS}}^{(1)}=\{A\in \Lcal_{\textrm{HS}}\mid [A,Q], [A,P] \in \Lcal_{\textrm{HS}}\}.
$$
On it, $\langle A,B\rangle$ is always finite and defines a sesquilinear form, linear in $B$ and anti-linear in $A$. 
To see the above mathematical precautions are necessary, consider a pure state $\rho=|\psi\rangle\langle\psi|$. Then
$
\Tr [\rho, Q]^2=-2\Delta Q^2,  \Tr[\rho, P]^2=-2\Delta P^2.
$
These traces are finite if and only if $\langle \psi|(Q^2+P^2)|\psi\rangle=2\langle \psi|a^\dagger a|\psi\rangle +1$ is finite. In other words, if and only if the mean photon number of the state $|\psi\rangle$ is finite. This is not a very restrictive condition, but it is essential, and it is satisfied by all states commonly considered in the literature. We remark that, similarly, the right hand side of~\eqref{eq:wigner_sobolev} is in general not finite for arbitrary operators $A$ and $B$, even if they are both trace class. Indeed, whereas then the Wigner function is square integrable, nothing guarantees its gradient is as well. Whence the need to restrict to the space $\Lcal_{\textrm{HS}}^{(1)}$ for which this is the case: assuming $[\rho, Q]$, respectively $[\rho, P]$, are trace class implies indeed $\partial_{\alpha_2} W$, respectively $\partial_{\alpha_1} W$  are square integrable.
We now show that  the above sesquilinear form is non-degenerate, in the sense that $\vertiii{A}=0$ implies $A=0$. As a result, it is an inner product and $\Lcal_{\textrm{HS}}^{(1)}$ is a pre-Hilbert space. With the usual abuse of notation, we will write $\Lcal_{\textrm{HS}}^{(1)}$ for its completion as a Hilbert space as well. We give two proofs. First note that $\vertiii{A}=0$ implies $[A,Q]=0=[A,P]$. Since the Weyl-Heisenberg group acts irreducibly, an operator that commutes with both $Q$ and $P$ must be a multiple of the identity. Since $A$ is trace class, this implies $A=0$.  Alternatively, take $A=B$ in~\eqref{eq:wigner_sobolev}, note that $\langle A, A\rangle=0$ implies $\nabla W_A=0$. Hence $W_A$ is a constant. Since $\int |W_A|^2(\alpha)\rd \alpha<+\infty$, $W_A=0$ and hence $A=0$. 

Note that the right hand side of~\eqref{eq:wigner_sobolev} defines an inner product on the space of functions of $\alpha\in\C$. It is well known as a homogeneous Sobolev inner product~\cite{bachda11} and the space of functions for which $\|\nabla W\|$ is finite is the homogeneous Sobolev space denoted by $\dot H^{1,2}(\C)$, important in the study of partial differential equations. One can therefore think of $\Lcal_{\textrm{HS}}^{(1)}$ as a quantum or non-commutative Sobolev space. 

We now show~\eqref{eq:Kmatrix}. We have
$$
\Tr[\rho,Q][Q,\rho]=\sum_i p_i^2\Tr\left[|i\rangle\langle i|, Q\right]\left[Q,|i\rangle\langle i|\right] +\sum_{i\not=j} p_ip_j\Tr[|i\rangle\langle i|, Q][Q,|j\rangle\langle j|]. 
$$
Since $\Tr\left[|i\rangle\langle i|, Q\right]\left[Q,|i\rangle\langle i|\right] =2\Delta Q_i^2$ and, for $i\not=j$,
$$
\Tr[|i\rangle\langle i|, Q][Q,|j\rangle\langle j|]=-2|\langle i|Q|j\rangle|^2,
$$
 the result follows.

Continuity is a desirable property for any non-classicality measure~\cite{ryspvo17}: small changes in the state should lead to small changes in its non-classicality.  Since the ordering sensitivity is a norm, it is automatically continuous and so is therefore the induced distance to the set $\tilde\Ccal$.

We finally establish some simple geometric properties of $\tilde\Ccal$ helping to locate it more precisely within the unit ball of $\Lcal_{\textrm{HS}}^{(1)}$, and that will allow us to prove~\eqref{eq:encadrement}. We have seen that $\tilde\rho\in \tilde\Ccal$ satisfies $\vertiii{\tilde\rho}=1$ iff and only if $\rho$ is a coherent state, and hence a pure state. So the pure classical states all lie on the unit sphere of $\Lcal_{\textrm{HS}}^{(1)}$. It is natural to wonder how far the $\tilde\rho\in\tilde\Ccal$, which lie inside the unit ball,  can be removed from its surface,  the unit sphere.   In that perspective, we point out that, using~\eqref{eq:Kmatrix} with the Fock basis, one easily sees that for a thermal state $\rho_{\mathrm{th}}$ with mean photon number $\langle n\rangle$, 
$
\Os(\rho_{\mathrm{th}})=(2\langle n\rangle +1)^{-1}.
$
Hence, as $\langle n\rangle \to+\infty$, $\Os(\rho_{\mathrm{th}})=\vertiii{\tilde\rho_{\mathrm{th}}}\to 0$. Hence the set $\tilde \Ccal$ contains elements arbitrarily close to the center of the unit ball, far from the surface of the sphere of radius one.
It is tempting to think of such states as very classical. Since large photon number corresponds to high temperature, this is in agreement again with basic physical intuition. We can now show~\eqref{eq:encadrement}. The lower bound is obvious. By the definition of $\Ncal(\rho)$ and the triangle inequality for norms 
we have
$
\Ncal(\rho)\leq \vertiii{\tilde \rho-\tilde \rho_{\mathrm{th}}}\leq \vertiii{\tilde\rho}+\vertiii{\tilde\rho_{\mathrm{th}}}.
$
Taking $\langle n\rangle$ to infinity, we find $\Ncal(\rho)\leq \vertiii{\tilde\rho}.$

\section{On the non-classicality of multi-component cat states}\label{s:appCS}
Multi-component cat states~\cite{hara13,hodekopa16} are superpositions of $N$ coherent states placed at the points $\alpha_m=\alpha_0\exp(-i\frac{2\pi}{N}m)$:
\begin{equation}
|\mathrm{c}_q\rangle =  \frac1{N\sqrt{\tilde g(q)}}\sum_{m=0}^{N-1}e^{i2\pi mq/N}|\alpha_0 e^{-i2\pi m/N}\rangle, \quad \tilde g(q)=\frac1{N}\sum_{m=0}^{N-1}g(m)e^{-i2\pi qm/N},\nonumber
\end{equation}
with
\begin{equation}
g(m)=\exp\left\{|\alpha_0|^2\left(e^{i2\pi m/N}-1\right)\right\}.
\nonumber
\end{equation}
All expressions only depend on the integer $q$ modulo $N\geq 2$ and $\langle c_q|c_{q'}\rangle=\delta_{q,q' \text{mod} N}$; $N=2$ yields even/odd ($q=0,1$) cat states. 
We first establish that the ordering sensitivity of these states tends to infinity with $\alpha_0$ at fixed $N$ proving their strong non-classicality.  We will then show that neither the Mandel parameter $Q_{\textrm{M}}$, nor the degree of squeezing $S$, defined below, nor the moment method~\cite{agta92} detect the non-classicality of the $N$ component cat states for large $|\alpha_0|$.

We have, for all $\ell\in\Z$: 
\begin{equation}\label{eq:RICSmoments}
a^\ell |c_q\rangle =\alpha_0^\ell \sqrt{\frac{\tilde g(q-\ell)}{\tilde g(q)}}|c_{q-\ell}\rangle,\quad
 \langle c_q|(a^\dagger)^\ell a^\ell |c_q\rangle=\frac{\tilde g(q-\ell)}{\tilde g(q)}|\alpha_0|^{2\ell}.
\end{equation}
Hence $\Os(N,q)=2\frac{\tilde g(q-1)}{\tilde g(q)}|\alpha_0|^{2}+1$ which tends to infinity with growing $|\alpha_0|$, proving the first assertion. 
We now turn to the moment method. It is shown in~\cite{agta92} that a sufficient condition for the non-classicality of a state $\rho$ is the negativity of one of the following determinants:
\begin{equation}
D_n=\left|
\begin{matrix}
1& m_1&\dots& m_{n-1} \\
m_1& m_2&\dots &m_n \\
\vdots&\vdots&\vdots&\vdots\\
m_{n-1} & m_n&\dots& m_{2n-2}
\end{matrix}
\right|,\quad \textrm{where}\quad m_\ell=\Tr\rho (a^\dagger)^\ell a^\ell.\nonumber
\end{equation}
Note that $D_2=Q_{\textrm{M}}$, the Mandel parameter.

In general, the computation of the $D_n$ becomes increasingly complex with growing $n$, since it requires the computation or measurement of higher order moments $m_\ell$. Here, using~\eqref{eq:RICSmoments} one finds, for general $\alpha_0, N, q, n$, 
\begin{equation}
D_n(q)=\frac{|\alpha_0|^{2n(n-1)}}{\tilde g(q)^n}
\textrm{det}\,\tilde G(q) ,\nonumber
\end{equation}
where
\begin{equation}\label{eq:Gdef}
\tilde G(q)=
\begin{pmatrix}
{\tilde g(q)}& {\tilde g(q-1)}&\dots& {\tilde g(q-(n-1))} \\
{\tilde g(q-1)}& {\tilde g(q-2)}&\dots &{\tilde g(q-(n-1)-1)} \\
\vdots&\vdots&\vdots&\vdots\\
{\tilde g(q-(n-1))} & {\tilde g(q-(n-1)-1)}&\dots& {\tilde g(q-2(n-1))}
\end{pmatrix}.
\end{equation}
Because of the $N$-periodicity of $\tilde g$, $D_n(q)=0$ for $n>N$. To compute $D_n$, we compute, for $m\not=0$ mod $N$,  
$$
|g(m)|=\exp\left(|\alpha_0|^2(\cos(\frac{2\pi m}{N})-1)\right)\leq \exp\left(-\frac{|\alpha_0|^2}{2}\eta\left(\frac{2\pi}{N}\right)^2\right),
$$
with $\eta=1-\frac1{12}\left(\frac{2\pi}{N}\right)^2$, since, for all $x\in\R$,
$
\cos(x)-1+\frac{x^2}{2}\leq \frac{x^4}{24}.
$
Hence
$$
\tilde g(q)=\frac1N +r(q),\quad\textrm{with}\quad |r(q)|\leq \overline r=\frac{N-1}{N}\exp\left(-\frac{|\alpha_0|^2}{2}\eta\left(\frac{2\pi}{N}\right)^2\right).
$$
Note that the estimate on the error term $r(q)$ does not depend on $q$. 
Introducing $\underline e=N^{-1/2}(1,1,\dots, 1)$, we have
$$
D_n=\frac{|\alpha_0|^{2n(n-1)}}{\tilde g(q)^n}\textrm{det}\left(\underline e^T\underline e +R(q)\right),
$$
where $R(q)$ is obtained from $r(q)$ in the same manner as $\tilde G(q)$ from $\tilde g(q)$ in~\eqref{eq:Gdef}.
It follows from the matrix determinant lemma that 
$
\textrm{det}(R(q)+\underline e^T\underline e)=\textrm{det}R(q)+ \underline e\textrm{adj}R(q) \underline e^T,
$
where adj$R(q)$ is the adjugate matrix of $R(q)$, meaning the transpose of the matrix of its cofactors. Hence
$$
|D_n|=\frac{|\alpha_0|^{2n(n-1)}}{\tilde g(q)^n}
\left|\textrm{det}R(q)+ \underline e\textrm{adj}R(q) \underline e^T\right|\leq \frac{|\alpha_0|^{2n(n-1)}}{\tilde g(q)^n}n!\overline r^{(n-1)}\left[\overline r+\frac{n}{N} \right].
$$
Therefore, for fixed $N,n,q$, the determinant $D_n$ tends very quickly to zero with growing $\alpha_0$.  Its negativity is therefore increasingly difficult to observe. In this sense, the moment method does not efficiently detect the growing non-classicality of the multi component cat states for growing $\alpha_0$.  In addition, observing $D_2$ closely as a function of $\alpha_0$, one observes it oscillates around zero, changing sign regularly as $\alpha_0$ grows.

For the degree of squeezing S~\cite{agta92}, defined for each $\varphi\in[0,2\pi[$ by 
$$
S=\langle:(a \exp(i\varphi) +a^\dagger \exp(-i\varphi))^2:\rangle -\langle (a \exp(i\varphi) +a^\dagger \exp(-i\varphi))\rangle^2
$$
we find, since $\langle a\rangle=0$, 
$
S=\langle a^2 \exp(2i\varphi) +(a^\dagger)^2 \exp(-2i\varphi)+ 2a^\dagger a \rangle.
$
Now
$
a^2 |c_q\rangle =\alpha_0^2 \sqrt{\tilde g(q-2)\tilde g(q)^{-1}}|c_{q-2}\rangle.
$
Hence, if $N\geq 3$, $\langle a^2\rangle=0$. One has then
$
S=2\langle c_q|a^\dagger a|c_q\rangle=2\frac{\tilde g(q-1)}{\tilde g(q)}|\alpha_0|^{2}>0.
$
Since negativity of $S$ is a witness of non-classicality, one concludes that the non-classicality of the multi component cat states is not detected by the degree of squeezing if $N\geq 3$. 
The case $N=2$ is slightly different. Then $a^2 |c_q\rangle=\alpha_0^2|c_q\rangle$ and, for $q=0,1$
$$
S(2,q)=2\alpha_0^2\left(\cos(2\varphi) + \frac{\tilde g(q-1)}{\tilde g(q)}\right), \quad
\tilde g(1)=1-\langle\alpha|-\alpha\rangle,\quad \tilde g(0)=1+\langle\alpha|-\alpha\rangle.
$$
So the even cat shows squeezing, the odd one does not. 

\section{Nonclassicality versus quantum macroscopicity}\label{s:appQM}
The Quantum Fisher Information (QFI) $\mathcal F(\rho, A)$ of the state $\rho$ for the observable $A$  is defined as~\cite{brca94}
\begin{equation*}
\mathcal F(\rho, A)=4\partial_x^2 D_B^2(\rho, \exp(-ixA)\rho\exp(ixA)_{|x=0},
\end{equation*}
where $D_B^2(\rho,\sigma)=2(1-F(\rho, \sigma))$ is the Bures distance and $F(\rho,\sigma)=\Tr \sqrt{\sqrt{\rho}\sigma\sqrt{\rho}}$ the fidelity between $\rho$ and $\sigma$.  Explicitly, 
\begin{equation}\label{eq:fisherexpression}
\mathcal F(\rho, A)=2\sum_{i,j}\frac{(p_i-p_j)^2}{p_i+p_j} |\langle i|A|j\rangle|^2=4\sum_{i<j}\frac{(p_i-p_j)^2}{p_i+p_j} |\langle i|A|j\rangle|^2,
\end{equation}
where $\rho=\sum_i p_i |i\rangle\langle i|$ is a spectral decomposition of $\rho$ and the sum is over $i,j$ for which $p_i+p_j>0$. It was proven in~\cite{tope13, yu13} that the QFI is equal to the convex roof of four times the variance of $A$  seen as a function on pure states. As pointed out in Section~\ref{s:QM}, the authors of~\cite{ousefrgias15, yave16} propose a measure $\MQFI(\rho)$ of quantum macroscopicity for the state $\rho$ of a bosonic field using the Fisher information as follows:
\begin{equation}
\MQFI(\rho)=\frac14\sup_\theta \mathcal F(\rho, Q_\theta),\quad Q_\theta=\cos\theta Q+\sin\theta P.\nonumber
\end{equation}
We now show equation~\eqref{eq:mqfiwitness}. Since $\mathcal F(\rho,Q_\theta)$ is convex, and  for a coherent state, $\Delta Q_\theta^2=\frac12$, one has, for any classical
$
\rho=\int P(\alpha) |\alpha\rangle\langle \alpha| \rd \alpha,
$
$$
\MQFI(\rho) \leq \int P(\alpha) \frac14\mathcal F(|\alpha\rangle\langle\alpha|, Q_\theta)\rd \alpha= \frac12.
$$
Consequently, if  $\MQFI(\rho)>\frac12$, then $\rho$ is nonclassical, proving the claim that $\MQFI$ is a nonclassicality witness. 

To further explore the relation between $\OS$ and $\MQFI$, we  compute $\MQFI$ for the various benchmark states studied in the main part of the paper. If $\rho$ is of the form $\rho=\sum_n p_n|n\rangle\langle n|$, where the $|n\rangle$ are the Fock states then
\begin{equation*}
\mathcal F(\rho,Q_\theta)=4\sum_{n<m} \frac{(p_n-p_m)^2}{p_n+p_{m}}|\langle n|Q|m\rangle|^2=
2\sum_{n}\frac{(p_n-p_{n+1})^2}{p_n+p_{n+1}}(n+1)
\end{equation*}
and
\begin{eqnarray*}
\frac12 \Tr [\rho, Q][Q,\rho]&=&\frac12 \Tr [\rho, P][P,\rho]=\sum_{n<m} (p_n-p_m)^2|\langle n|Q|m\rangle|^2\nonumber\\
&=&\frac12\sum_{n}(p_n-p_{n+1})^2(n+1).
\end{eqnarray*}
First, consider states $\rho_{N, M}$, with $p_n=M^{-1}$ for all $N+1\leq n\leq N+M$ and $p_n=0$ otherwise, for some $N\geq 0,M>0$, two integers. Then
$$
S_0(\rho)=1+2\frac{N+1}{M}>1, \quad \MQFI(\rho)=\frac14\mathcal F(\rho, Q)=\frac12+\frac{N+1}{M}>\frac12
$$ 
so that
$
\MQFI(\rho)=\frac12 \Os(\rho).
$
So, for these states, the ordering sensitivity introduced here and the quantum macroscopicity measure of~\cite{ousefrgias15, yave16} coincide up to a global normalization. Note that these states are non-classical and both quantities detect this. Their Mandel parameter, on the other hand, one finds
$$
Q_{\textrm{M}}=(\frac1{12}M(M-6)-\frac7{12}-N)\langle a^\dagger a\rangle^{-1},\quad \langle a^\dagger a\rangle = \frac12(M+1)+N.
$$
This is positive for $M$ sufficiently large and therefore does not detect their nonclassicality.

For thermal states $\rho_{\textrm{th}}$, where $p_n=\mathcal N_\lambda \exp(-\lambda n)$ one has similarly
$$
S_0(\rho_{\textrm{th}})=\Tr\rho_{\textrm{th}}^2=\frac12 \mathcal F(\rho_{\textrm{th}},Q)=2\MQFI(\rho_{\textrm{th}}).
$$
Note that, since the thermal states are classical, one has indeed $\OS(\rho_{\textrm{th}})<1$. 

For the Truncated Thermal States introduced in~\cite{le95}, given by
$
\rho_{\textrm{TTS}}=(\exp(\beta)-1)\sum_{n=1}^{+\infty} \exp(-\beta n) |n\rangle\langle n|, 
$
one has
$
Q_{\textrm{M}}(\rho_{\textrm{TTS}})=(\exp(\beta)-1)^{-1}-1=\langle\hat n\rangle_\beta-2,
$
and
$$
\OS(\rho_{\textrm{TTS}})=(1-\e^{-\beta})\left(2+\frac1{1+\e^{-\beta}}\right),\quad
\MQFI(\rho_{\textrm{TTS}})=\frac32\frac{1-\e^{-\beta}}{1+\e^{-\beta}}.
$$
It follows the nonclassicality of these states is detected by $Q_{\textrm{M}}$ and by $\MQFI$ provided $\beta>\ln 2\simeq 0.6931$  and by $\OS$ for $\beta>\ln\frac{2}{\sqrt 5-1}\simeq 0.4812$. Consequently,  the ordering sensitivity outperforms the two other witnesses for these states.

In the previous families of examples, when there are differences between the behaviour of the nonclassicality measure $\Ncal$ and the quantum macroscopicity measure $\MQFI$, they occur at low levels of nonclassicality, whereas the asymptotically large values tend to agree. We now give a simple example where the two differ considerably for large values. For that purpose, let us consider 
$
\rho_k=(1-\epsilon_k)|0\rangle\langle 0| +\epsilon_k |k\rangle\langle k|.
$
Then 
$$
\OS(\rho_k)=1+2\frac{k\epsilon_k^2}{(1-\epsilon_k)^2+\epsilon_k^2}\quad\textrm{and}\quad
\MQFI(\rho)=\frac12+ k\epsilon_k.
$$
Taking $\epsilon_k=M_*k^{-1}$, one has $\MQFI(\rho_k)=\frac12 +M_*$ and $\OS(\rho_k)\to 1$ as $k\to+\infty$. In fact 
$
\Ncal(\rho_k)\leq |||\tilde\rho_k-|0\rangle\langle 0||||\to 0.
$
This example shows that $\MQFI$ can be arbitrarily large, while the nonclassicality $\Ncal$ is arbitrarily small. This means that, whereas $\MQFI$ can act as  a nonclassicality witness, it cannot serve as a nonclassicality measure.  Consequently, whenever one finds a state with a large value of the quantum macroscopicity measure $\MQFI$, one cannot conclude it has a large nonclassicality or ordering sensitivity. In the last example, the large value of $\MQFI$ must be attributed to another property of the state than its ordering sensitivity.

As a final example, we consider squeezed thermal states. Those are defined as
$
\rho_{\textrm{sqth}}=S(z)\rho_{\textrm{th}}S(z)^\dagger,
$
where $z=r\exp(i\varphi)$ is the squeezing parameter and $S(z)=\exp\left\{\frac{z^*}{2}a^{2}-\frac{z}{2} a^{\dagger2}\right\}$. Note that they are not rotationally invariant. Their eigenstates are $S(z)|n\rangle$ and using~\eqref{eq:fisherexpression}, one finds
$
\mathcal F(\rho_{\textrm{sqth}}, Q_\theta)= |C_\theta|^2 \frac2{2\langle n\rangle_{\textrm{th}}+1}
$
where $C_\theta=\cosh r -\exp(2i\theta -i\varphi)\sinh r$.
Hence
$
\MQFI(\rho_{\textrm{sqth}})=\MQFI(\rho_{\textrm{th}})\exp(2r)=\frac12\frac1{2\langle n\rangle_{\textrm{th}}+1}\exp(2r).
$ It follows that $\MQFI(\rho_{\textrm{sqth}})>1/2$ iff $r>\frac12\ln (2\langle n\rangle_{\textrm{th}}+1)$. It is known~\cite{kidekn89} that this condition is necessary and sufficient for the nonclassicality of squeezed thermal states so that $\MQFI$ detects this property optimally for those states. For $\OS$, one finds, on the other hand
$
\OS(\rho_{\textrm{sqth}})=\frac1{2\langle n\rangle_{\textrm{th}}+1}\cosh(2r). 
$
The condition $\OS(\rho_{\textrm{sqth}})>1$ yields a condition on $r$ for nonclassicality, which is however suboptimal in this case. So for this example, the quantum macroscopicity $\MQFI$ is a better nonclassicality witness than the ordering sensitivity $\OS$. Note however this. 
From~\eqref{eq:encadrement}, we see that, asymptotically for large $r$, $\Ncal(\rho_{\textrm{sqth}})\sim \exp(r)$: the squeezed thermal states are exponentially far from classical in the squeezing parameter $r$. 
From the large value of $\MQFI$, such information cannot be inferred a priori, as the previous example shows. 


\end{document}